%% file: case.tex
\documentclass[10pt,twocolumn]{article}
\usepackage{fullpage}
\usepackage{epsfig}
\usepackage{times}
\usepackage{paralist}
\usepackage{barenote}

\long\def\com#1{}
\begin{document}

\title{{\bf Unmanaged Internet Protocol} \\
	Taming the Edge Network Management Crisis}
\author{Bryan Ford \\ Massachusetts Institute of Technology}
\date{}

\maketitle

\barenote{
This research was conducted as part of the IRIS project
({\tt http://project-iris.net/}),
supported by the National Science Foundation
under Cooperative Agreement No. ANI-0225660.}

\input{abs}
\input{intro}

\input{motiv}

\input{arch}
\input{route}
\input{deploy}
\input{related}

\input{conc}

\com{
Outline:

What do networked edge devices do well?
- make it easy to attach an edge device to the network for use as a client

What do networked edge devices not do well?
- address and connect with other edge devices, especially when they move
- communicate with other nearby edge devices while Internet is unavailable
- be able to attach at different points without reconfiguring each time
- be able to attach at multiple points and use both connections at once,
  or even just be able to fail-over from one to the other.
}

\begin{small}
\bibliography{case}
\bibliographystyle{plain}
\end{small}

\end{document}

%% file: abs.tex
\begin{abstract}

\begin{small}
Though appropriate for core Internet infrastructure,
the Internet Protocol is unsuited to routing
within and between emerging ad-hoc edge networks
due to its dependence on hierarchical,
administratively assigned addresses.
Existing ad-hoc routing protocols address the management problem
but do not scale to Internet-wide networks.
The promise of ubiquitous network computing cannot be fulfilled
until we develop an {\em Unmanaged Internet Protocol} (UIP),
a scalable routing protocol that manages itself automatically.
UIP must route within and between constantly changing edge networks
potentially containing millions or billions of nodes,
and must still function
within edge networks disconnected from the main Internet,
all without imposing
the administrative burden of hierarchical address assignment.
Such a protocol appears challenging but feasible.
We propose an architecture
based on self-certifying, cryptographic node identities
and a routing algorithm adapted from distributed hash tables.

\com{
Unmanaged Internet Protocol (UIP)
provides scalable, fully self-managing routing
re-implements the original ARPAnet vision
of scalable, robust, any-to-any communication between all participating hosts,
on the discontinuous and rapidly changing fabric of the modern Internet.
In essence,
UIP stitches together the currently fragmented Internet
into a single network
with uniform addressing and universal connectivity.
UIP provides decentralized, management-free routing
among hosts on the existing IPv4 and IPv6 Internets,
hosts on private networks
behind firewalls and network address translators,
mobile hosts with ephemeral IP addresses,
and hosts on experimental or ad-hoc wireless networks
with no IP-based connectivity.
Applications address other UIP nodes
using self-certifying cryptographic identities
that can be created by anyone,
require no centralized administration,
and remain valid as long as desired even as nodes move.
All communication between UIP nodes
is integrity- and privacy-protected by default,
making application-specific security protocols unnecessary.
}

\com{
Users can register their UIP identifiers
with public name services such as DNS for convenience,
or they can keep their node identifiers anonymous
to permit secure remote access
without exposing information about affiliations or private networks.
UIP directly supports multi-homing, host mobility, disconnected operation,
and remote access through legacy NATs and firewalls.
Application changes required to support UIP
are comparable to those required for IPv6.
}
\end{small}

\end{abstract}

%% file: intro.tex
\section{Introduction}

The promise of ubiquitous computing
is that people will soon routinely own many ``smart'' networked devices,
some mobile, others perhaps built into their homes and offices,
and all of which they can access and control from any location
so long as appropriate security precautions are taken.
Before we can expect ordinary, non-technical people
to adopt this vision, however,
the ad-hoc {\em edge networks} in which these devices live
must be able to manage themselves.
Each device must be able to find and communicate with its peers---%
whether connected directly,
indirectly over a local-area network,
or remotely across a long distance via the Internet---%
with no special configuration or other technical effort on the part of the user.

The current Internet Protocol
is unsuited to this purpose.
IPv4 and IPv6,
with their accompanying routing, naming, and management protocols,
have evolved around the requirements
of {\em core} network infrastructure:
corporate, academic, and government networks deployed and managed
by skilled network administrators.
IP's hierarchical address architecture in particular
is fundamentally dependent on skilled network management.
Current ad-hoc networking protocols by themselves are not sufficient either,
because they are only scalable to local-area network sizes
of a few hundreds or thousands of nodes.

\begin{figure}[t]
\centerline{\epsfig{file=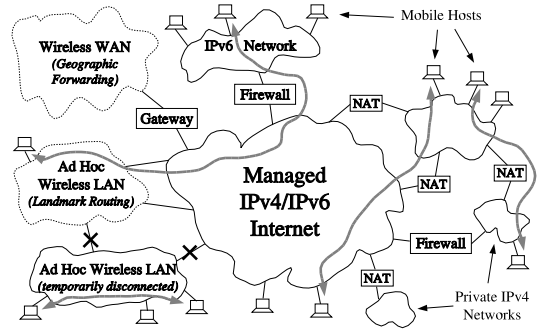, scale=0.35}}
\caption{Today's Internetworking Challenges}
\label{fig-blobs}
\end{figure}

To achieve ubiquitous network computing,
we need an {\em Unmanaged Internet Protocol}, or UIP, 
that combines the self-management of ad-hoc networks
with the scalability of IP.
As illustrated in Figure~\ref{fig-blobs},
achieving this goal in today's chaotic mix of networking technologies 
also means routing traffic automatically and securely through NATs,
and transparently bridging IPv4, IPv6,
and other address domains.
We propose an architecture based on
{\em scalable identity-based routing},
or routing based on topology-independent node identifiers.
While more difficult than routing
over topology-dependent addresses such as IP addresses,
there is evidence that scalable identity-based routing
is possible and practical.

This position paper is organized as follows.
Section~\ref{sec-motiv} lays out the motivation for UIP
and the inadequacies of current solutions.
Section~\ref{sec-arch} proposes and outlines
an identity-based UIP routing architecture,
and Section~\ref{sec-status} describes implementation status and deployment.
Section~\ref{sec-related} summarizes related work,
and Section~\ref{sec-conc} concludes.

%% file: motiv.tex
\section{Motivation for UIP}
\label{sec-motiv}

The original ARPAnet vision
was to enable computer users
to communicate and share resources with users of any other
connected computer~\cite{licklider68computer, roberts70computer}.
This vision has evolved into the modern Internet Protocol,
whose purpose is to implement any-to-any connectivity between hosts,
whether connected directly
or indirectly via paths crossing many administrative domains.
While physical and link-layer technologies such as Ethernet
provide low-level building blocks for communication,
and higher-level protocols
enable applications and users to take advantage of the network,
{\em interoperable end-to-end connectivity} via IP
remains the Internet's central focus.

Technical, social, and economic pressures
have hindered the achievement of this vision, however.
The protocols underlying the Internet were designed
by technically savvy individuals
who understand how networks work
but often do not understand how {\em non-technical users} work.
As a result many aspects of network operation
still require careful and skilled management.
We desire and increasingly expect
that {\em everyone} should be able to
use the Internet and deploy networked devices,
and strong economic incentives exist
for businesses to sell Internet-enabled hardware and Internet-based services
to technically unsophisticated users.
Since businesses seek
lowest-cost paths to profitable solutions,
the commercial Internet has evolved---%
via a chaotic series of hacks and extensions---%
into a system geared toward particular usage patterns
that facilitate business opportunities,
often at the expense of interoperability
and general end-to-end connectivity.

\subsection{The Edge Network Management Crisis}

An unsophisticated user can now buy a computer,
connect it to the Internet, and use it for browsing the Web,
reading E-mail, and shopping on-line.
Users who are a bit more adventurous but still relatively non-technical
may set up a small home network
and surf the Web from several computers at once.
But consider the following scenario:
\begin{enumerate}
\item	Joe User is working at home on his laptop.
	He has remote shell and database access sessions open,
	through his WiFi home network,
	to his desktop PC and to a machine at his workplace.
\item	Joe's friend Jim calls and invites him over.
	Joe puts his laptop into sleep mode and hops into his car.
\item	Joe stops for a bite to eat on the way to Jim's,
	and scribbles some notes on his PDA in the restaurant.
\item	Upon returning to his car,
	Joe tries to synchronize his PDA with his laptop,
	but discovers they won't talk to each other
	even though they're both WiFi-enabled and are at most a foot apart.
	Being unfamiliar with the technical details of IP networks,
	he doesn't realize that this is because
	(a) the WiFi adapters are in infrastructure
	rather than ad-hoc mode,
	and (b) even if they could communicate at the link layer,
	neither machine would be able to get an IP address
	because there's no DHCP~\cite{rfc2131} server nearby.
\item	Joe arrives at Jim's place,
	and the two brainstorm about their project at work.
	Joe takes out his laptop and wakes it up.
	Since Jim also has an Internet-connected WiFi home network,
	Joe hopes to use Jim's Internet access
	and resume his existing application sessions
	to his desktop PCs at home and work.
	Again Joe is disappointed.
	After figuring out that he has to remove and re-insert
	his laptop's WiFi card in order to get it
	to recognize Jim's network at all,
	Joe's application sessions are gone.
	He does not realize that
	moving to a new attachment point changed his IP address,
	breaking his existing TCP connections.
\item	Joe tries to re-start his application sessions,
	but finds that he cannot even locate
	let alone connect with his desktop PC at home.
	He doesn't realize that this is because
	(a) his ISP did not give him a permanent IP address
	useable for connecting remotely to his home network,
	and (b) even with a permanent IP address,
	his desktop PC would still be inaccessible
	because it is behind a network address translator~\cite{rfc3022}.
\com{	awww, nobody appreciates my humor...
\item	Joe becomes disgusted with technology,
	throws out his computers, says goodbye to Jim,
	and moves to the countryside to grow grapes.
}
\end{enumerate}

Joe's na\"ive expectations of his networked devices
are not fundamentally unreasonable,
and all of the problems above are solvable with current technology.
Joe could in theory:
(a) configure his home NAT to assign fixed site-local IP addresses
to his desktop and laptop PCs at home;
(b) configure his NAT to open the appropriate external ports for remote access
and forward incoming connections on those ports to his desktop PC;
(c) register for a global DNS host name
with a Dynamic DNS~\cite{rfc2136} service provider;
(d) set up his desktop PC to update this DNS name periodically
with the dynamic IP address his ISP assigns to his home NAT;
(e) set up Mobile IP~\cite{rfc3344}
so that his desktop PC at home will intercept packets
destined for his laptop's ``home'' IP address,
and tunnel them to his laptop at its actual connection point
while connected elsewhere;
(f) run daemons on his laptop and PDA
that detect when no infrastructure-mode WiFi access point or DHCP service
is available,
and automatically switch into ad-hoc mode
using a routing protocol such as AODV~\cite{perkins99adhoc}.

Only the most dedicated, desperate, or geeky
will go to this trouble, however.
To most users,
having a ``working'' network
means being able to get to Google, CNN, and Amazon.com.
Any ``ubiquitous'' connectivity
outside this commercial client/server straitjacket
is fickle, unreliable, and management-intensive
if available at all.

\subsection{IP Networks Require Management}

The scalability and efficiency of the current Internet Protocol
relies on Classless Inter-Domain Routing (CIDR)~\cite{rfc1518},
in which network nodes are assigned addresses
whose hierarchical structure reflects the routing topology.
BGP routers take advantage of the hierarchical structure of IP addresses,
aggregating information about distant nodes and networks
sharing a common address prefix into a single routing table entry~\cite{rfc1771}.

While this hierarchical address assignment scheme
makes the core Internet infrastructure efficient and scalable,
it is precisely this address assignment scheme
that makes edge networks brittle and difficult to manage.
Whenever a node moves or its surrounding network is renumbered,
the node's IP address must change.
Statically configuring and maintaining
the IP addresses of many nodes is challenging
even for technically competent network administrators,
leading to organizational resistance
against IP address renumbering~\cite{rfc2101}.
Dynamic address assignment
transfers administrative responsibility
from edge nodes to DHCP servers,
at the expense of making edge nodes unable to communicate at all
without access to a DHCP server.
Workarounds in which nodes choose their own local IP addresses
after failing to contact a DHCP server~\cite{microsoft98apipa}
are slow, unreliable,
and at best allow nodes to communicate only with immediate link-neighbors
while disconnected from the main Internet.
These issues will persist even into a future IPv6 world
in which there are ``enough'' IP addresses for everyone
and network address translators do not exist,
because the basic address architecture remains the same in IPv6.

\com{
DHCP~\cite{rfc2131} makes it easy to attach new nodes to a network,
but renders IP addresses useless as stable node identifiers.
Dynamic DNS~\cite{rfc2136} and other solutions
to the resulting ``node identity crisis''
require setup and maintenance and remain sparsely used.
Reliance on DHCP also cripples nodes
that become temporarily disconnected from the Internet~\cite{microsoft98apipa}.
Mobility and multihoming
violate the Internet's hierarchical addressing and routing model~\cite{rfc1518},
%\cite{rfc1518, rfc1519, rfc1887}
leading to complex extensions~\cite{rfc3344, rfc3178}
that demand additional care and feeding.
Firewalls and network address translators~\cite{rfc3022}
create addressing and routing discontinuities in the Internet,
making remote access and peer-to-peer communication difficult~\cite{rfc3027}.
Since secure communication protocols
were not in the original Internet design,
they evolved late, inconsistently, and in application-specific niches:
SSL/TLS~\cite{rfc2246} for Web access,
SSH~\cite{ylonen02ssh} for remote terminal access,
IPSEC~\cite{rfc2401} for virtual private networks.
Ad-hoc wireless networking
protocols~\cite{tsuchiya88landmark, ko98locationaided}
show promise but require new address architectures
incompatible with those of the existing Internet.
}%com

\subsection{Ad-Hoc Networks Do Not Scale}

Classic
distance-vector~\cite{ford62flows, rfc1058}
and link-state~\cite{rfc1247} routing protocols,
as well as ad-hoc routing variants
such as DSR~\cite{johnson94routing}
and AODV~\cite{perkins99adhoc},
require every node to store and regularly exchange
information about {\em every} other node in the network.
This linear per-node storage and/or bandwidth overhead
limits the scalability of these protocols
to a few hundreds or thousands of nodes.
While ad-hoc protocols can be used to route within a particular IP subnet,
this subnet must be centrally allocated and managed
in order to be globally routable on the Internet,
and all participating nodes must be assigned to that subnet.
Configuring each node statically is tedious and inconvenient,
while using DHCP again makes the nodes unable to communicate with each other
while out of range of a DHCP server.

To fulfill the promise of ubiquitous networking,
an edge network routing protocol
must be self-managing not only on a {\em local} scale,
but also on a {\em global} scale.
We need an ad-hoc routing protocol that can
seamlessly route packets throughout an {\em Internet-wide federation}
of ad-hoc edge networks,
consisting of potentially millions or billions of edge nodes
that frequently hop from one edge network to another.
This protocol must still provide reliable ad-hoc routing
within edge networks
that are temporarily or permanently disconnected from the Internet.
This is the purpose of {\em Unmanaged Internet Protocol}, or UIP.

\com{
We are developing an experimental protocol for this purpose,
whose architecture we outline in the remainder of this paper.
The key feature of our architecture
is that it implements scalable routing
on topology-independent {\em node identities}
rather than on topologically-structured node addresses,
allowing nodes to choose their own unique identifiers cryptographically.
and to retain their identifiers as the network changes.
While our specific UIP architecture
is not necessarily the ``right'' or ``best'' possible one
(it probably isn't),
the important point
is that {\em some} network-layer architecture with similar properties
is essential to bringing about the reality
of ubiquitous network computing,
and we hope to shine a light in that direction.
}

\com{
XXX Define purpose more clearly:
allow all nodes to have identifiers that
	- require no manual configuration
	- are valid and stable
		for as long as the node wants them to be
	- can be used to initiate a secure connection
		from any other node anywhere else on the network.

XXX More clearly answer the question:
"we have addresses and names; why do we also need identifiers?"
}

%% file: arch.tex
\section{Proposed UIP Architecture}
\label{sec-arch}

\begin{figure}
\centerline{\epsfig{file=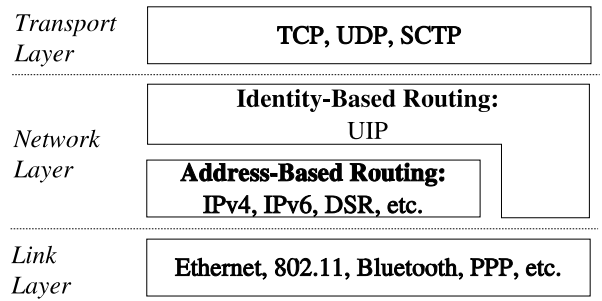, scale=0.4}}
\caption{UIP in the Internet Protocol Architecture}
\label{fig-layers}
\end{figure}

Since IP does an excellent job of routing packets efficiently
through the managed core Internet infrastructure,
we intend UIP not to replace IP but to run on top of it,
as a new network layer component (Figure~\ref{fig-layers}).
In our proposed architecture,
UIP appears to upper-level transport and application protocols
as a new address/protocol family,
much like IPv6 does now.

\subsection{Node Identities}

To refer to other UIP nodes,
applications use self-certifying cryptographic {\em identifiers}
that are stable over time and independent of network topology.
All connections between UIP nodes
are privacy- and integrity-protected by default,
as in IPSEC~\cite{rfc2401}.
A node's UIP identifier is a hash of the node's public key,
making identifiers {\em self-certifying},
like Moskowitz's host identities~\cite{moskowitz03hip-arch}
or SFS pathnames~\cite{mazieres99separating}.
Cryptographic identities
provide several properties
crucial to achieving robust connectivity
in future ubiquitous networking environments:

\begin{itemize}
\item	Any node can create a globally unique UIP identifier
	at any time without reference to central authorities.
	The identifier's uniqueness depends only on
	the strength of the cryptographic hash used to create it.
\item	A node's identifier remains valid as long as desired.
	Security practice may limit
	the lifetimes of long-term keys and node identities
	to a few years,
	but since this is the useful lifetime of most PCs,
	many nodes may never have to change identifiers.
\item	Since a node identifier contains no topology information,
	the node can retain its identity when it moves
	or the surrounding network topology changes.
\item	A node can cryptographically prove ownership of an identifier
	using the associated private key,
	preventing an attacker from stealing its identity.
\item	A node can have multiple identities simultaneously,
	representing distinct services or ``virtual hosts''
	on one physical machine for example.
\item	The network layer does not depend on
	centralized public key infrastructure (PKI).
	Higher layers may use PKI to map convenient names to node identifiers,
	but given a node identifier,
	finding and connecting securely to that node is fully decentralized.
\end{itemize}

\com{	not highly relevant, potentially confusing, and dispensable...

\begin{figure}
\centerline{\epsfig{file=naming.eps, scale=0.4}}
\caption{UIP in the Internet Naming Architecture}
\label{fig-naming}
\end{figure}

Figure~\ref{fig-naming} illustrates
how UIP fits into the Internet naming scheme.
Application protocols can use UIP identifiers directly,
embedding them into URLs for example
to form secure, self-authenticating object or service location strings
similar to self-certifying pathnames in SFS~\cite{mazieres99separating}.
Alternatively, applications can use DNS
to look up a node's UIP identifier from a human-readable domain name,
in the same way that they currently resolve IP addresses.
Since UIP identifiers are stable across topology changes, however,
DNS tables need to be changed much less frequently.
}%com

%com

%% file: route.tex
\subsection{UIP Routing}
\label{sec-route}

UIP's primary technical challenge
is to forward traffic from any node to any other
in an Internet-scale network,
without the help of hierarchically structured node addresses.
Since UIP node identifiers are unrelated to network topology,
they have no locality properties
routers can use to aggregate routing information about distant nodes.
Requiring every node to store and propagate routing information
about every other node in an Internet-scale network
may arguably be viable for a desktop PC
with a high-speed Internet connection,
but is definitely impractical for small, low-power devices such as PDAs.

\subsubsection{Approaches to Identity-Based Routing}

Bellman-Ford and similar routing algorithms find {\em optimal} routes,
based on either hop count or some per-link cost metric.
We do not need optimal routing, however:
in practice it suffices to find {\em reasonably efficient} routes.
The routes that BGP finds
are probably less than optimal already,
due to the difficulty of supporting site multihoming
in the Internet's hierarchical address model~\cite{rfc3582},
and the lack of incentive for ISPs
to reveal all of their peering relationships
in their BGP advertisements.

Route efficiency is usually measured in terms of {\em stretch}:
the length of the route discovered by the protocol
over the length of the best possible route.
Algorithms now exist
that can route through an $N$-node network with arbitrary node labels,
using $\tilde{O}(\sqrt{N})$ bits of routing table state per node,
and a small constant maximum stretch~\cite{arias03compact}.

In practice,
we do not even necessarily demand that {\em all} nodes
have sublinear storage requirements.
We might accept a routing protocol
that has sublinear overhead on {\em most} nodes,
but requires {\em a few} nodes present in the network
to have $\Omega(N)$ storage and/or bandwidth.
It is critical to our ubiquitous networking goals, however,
that these ``supernodes'' do not need to be hard-wired as such.
{\em Any} node must be able to take on that role dynamically
whenever supernodes are needed
and the surrounding network is small enough.
While Joe's laptop and PDA are connected to the Internet, for example,
their ability to route to distant edge nodes
might depend on a massive central server somewhere
that continuously maintains a complete map of the Internet.
If Joe takes several network devices
with him into the mountains
where there is no Internet access, however,
each device must be able
to take on super\-node responsibilities as necessary
to direct traffic within any smaller ad-hoc network he may form.
Joe's laptop and other small devices never need to map the entire Internet,
but only the smaller edge networks Joe may participate in
while disconnected from the Internet.

\subsubsection{Converting DHTs into Routing Algorithms}

\begin{figure}
\centerline{\epsfig{file=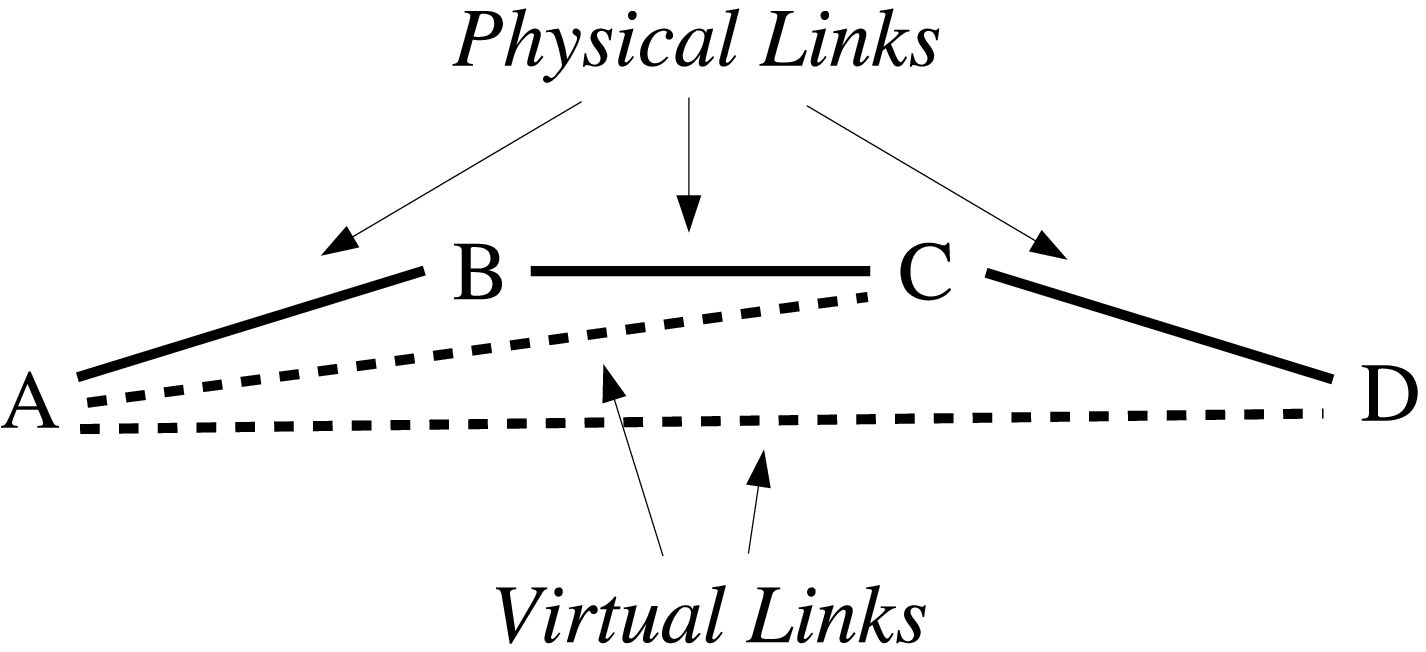, scale=0.4}}
\caption{Forwarding via Virtual Links}
\label{fig-vlinks}
\end{figure}

We are experimenting with a scalable routing protocol for UIP
derived from the Kademlia
distributed hash table (DHT)~\cite{maymounkov02kademlia}.
This protocol, detailed elsewhere~\cite{ford03scalable},
empirically achieves $O(log~N)$
storage and maintenance overhead per node
and an average stretch of 2 on simulated networks.
We have not yet formalized the algorithm
or derived theoretical performance properties, however.

DHTs normally implement only indexing and lookup,
relying on underlying protocols to provide connectivity
between all participants.
Our protocol extends Kademlia to function over any topology
through the use of recursive source routes, or {\em virtual links}.
Once any two nodes have connected and established a ``neighbor'' relationship,
ether node can use that relationship
to build further neighbor relationships recursively,
covering longer topological distances.
In Figure~\ref{fig-vlinks}, for example,
node A uses physical links AB and BC to build a virtual link AC,
thereby establishing a neighbor relationship with C.
Node A then uses virtual link AC
to build a recursive virtual link to C's physical neighbor D.
Node D can now communicate with node A via its physical link to C
and C's virtual link to A,
without having to know the details of the path between C and A.
Node D might not know about the existence of node B at all.
In this way the routing protocol
abstracts the details of routing at different levels,
achieving an effect analogous to IP address aggregation
without actually depending on hierarchically assigned addresses.

A node $n$ sorts its neighbors $n_i$ into {\em buckets},
according to the longest common prefix length in bits
between $n$'s and $n_i$'s identifier.
To join the network,
$n$ needs a physical link to some existing node $n_1$.
Node $n$ searches $n_1$'s neighbor table for another node $n_2$
with a longer common prefix,
and builds a virtual link through $n_1$ to $n_2$.
This process continues
until $n$ finds the node with the closest identifier to its own.
If a suitable connectivity invariant is maintained,
every node in the resulting structure
can find and build a forwarding path to any other node on demand.

%com

%% file: deploy.tex
\section{Implementation and Deployment}
\label{sec-status}

A UIP prototype is under development,
which we look forward to using and evaluating shortly.
For portability,
the prototype runs as an application-level daemon,
communicating with other nodes primarily over UDP.
The daemon can also directly utilize the link layers of some systems
if it has appropriate privileges.
Multiple local applications can share a single UIP daemon,
interacting through a proxy library
that exports a standard sockets-based interface.
To simplify initial deployment of UIP,
applications without special privileges
can be bundled with their own UIP daemon,
which the application starts automatically
if a systemwide daemon is not available.

An immediate benefit of UIP is that it allows applications
to establish secure, peer-to-peer connections
through NAT and firewall barriers without special effort,
as long as there are some widely-accessible UIP nodes on the Internet
through which the daemon can forward traffic if necessary.
The UIP daemon implements
UDP hole punching~\cite{ford03network}
as an application-transparent optimization,
utilizing widely-accessible UIP nodes as ``introducers''
to establish direct IP-based peer-to-peer communication paths
across many NATs and firewalls.
The daemon falls back to explicit forwarding whenever hole punching fails,
ensuring maximum robustness.

%com

%% file: related.tex
\section{Related Work}
\label{sec-related}

UIP is a fusion of ideas from many projects.
Like Resilient Overlay Networks~\cite{andersen01resilient},
UIP introduces a routing layer above IP
that can route around discontinuities and failures in the Internet,
but UIP seeks to be scalable and self-managing as well as resilient.
Ad-hoc routing protocols
such as DSR~\cite{johnson94routing}
address the management problem at the local level
but are not scalable to Internet-wide ad-hoc networks.
Landmark~\cite{tsuchiya88landmark}
and AODV~\cite{perkins99adhoc}
offer scalability under localized traffic patterns,
but not under the global traffic patterns of the Internet.

UIP node identities are similar
to those of Moskowitz's proposed
Host Identity Protocol~\cite{moskowitz03hip-arch},
but UIP uses identities for routing as well as authentication.
The Internet Indirection Infrastructure ($i3$)~\cite{stoica02internet}
provides location-independent host identities,
multicast and anycast communication,
and NAT traversal,
but does not implement a general-purpose routing protocol
that can function independently from the Internet.

Many Internet host mobility solutions have been proposed.
Higher-level naming systems
can provide applications independence
from their host's IP address~\cite{rfc2136,adjie-winoto99ins,snoeren00endtoend},
at the cost of tying applications to a particular naming scheme
and making it difficult to maintain connections
as the host moves~\cite{snoeren01reconsidering}.
Mobile IP~\cite{rfc3344} allows a mobile host to roam
without breaking outstanding TCP connections or UDP bindings,
but requires each mobile host to have a stable ``home'' IP address
through which packets are tunneled.
A similar illusion of a static IP address can be achieved
with IP multicast~\cite{mysore97multicasting, helmy00multicastbased}.
\com{
All of these mechanisms operate
in the context of the traditional IP addressing model,
and therefore may be easier to deploy in the short term than UIP.
On the other hand,
these existing mechanisms are complex,
require considerable management and configuration,
introduce substantial security implications,
and address only the mobility issue
and not the more general problem of uniform addressing and connectivity.
}
%UIP does not need any special mechanism to support mobility, however,
%because its node identities are location-independent
%and its routing protocol adapts to changing network topologies.

Work on peer-to-peer connectivity through firewalls and NATs~\cite{rfc3027}
has led to various special-purpose
protocols~\cite{rfc1928, rfc3489, upnp01igd}.
UDP hole punching~\cite{ford03network}
allows peer-to-peer connectivity through many NATs, but not all,
without the use of explicit proxy protocols.
Name-based routing~\cite{gritter01content}
offers more general bridging of IP address domains,
but its ties to the management-heavy domain name system
make it unsuitable for ad-hoc networks.

% XXX JXTA

\com{ XXX Higher-level solution being explored:
	Link-Local Multicast Name Resolution
	http://www.ietf.org/internet-drafts/draft-ietf-dnsext-mdns-16.txt
}

%% file: conc.tex
\section{Conclusion}
\label{sec-conc}

Ubiquitous network computing
will require an ad-hoc routing protocol
that can not only route autonomously
within small edge networks of hundreds or thousands of nodes,
but can seamlessly route
among a large Internet-connected {\em federation} of edge networks.
The traditional solution to scalability,
hierarchical address assignment,
is unsuited to edge networks
due to its management costs.
Scalable identity-based routing protocols appear feasible
and represent a promising research direction,
though many practical technical problems remain unsolved.
Besides providing a key building block for ubiquitous network computing,
scalable identity-based routing may also help
address the more immediate problems of Internet host mobility,
NAT traversal, and bridging between IPv4, IPv6, and other address domains.

\subsection*{Acknowledgments}

I wish to thank my advisor Frans Kaashoek,
my colleagues Dave Andersen and Chris Lesniewski-Laas,
Prof.\ David Karger,
and the HotNets reviewers
for many helpful comments and suggestions.